\documentclass[conference]{IEEEtran}
\IEEEoverridecommandlockouts

\usepackage{cite}
\usepackage{amsmath,amssymb,amsfonts}
\usepackage{algorithmic}
\usepackage{graphicx}
\usepackage{textcomp}
\usepackage{xcolor}
\usepackage{multicol}
\usepackage{array, multirow}
\def\BibTeX{{\rm B\kern-.05em{\sc i\kern-.025em b}\kern-.08em
    T\kern-.1667em\lower.7ex\hbox{E}\kern-.125emX}}
\begin{document}

\title{AAD-DCE: An Aggregated Multimodal Attention Mechanism for Early and Late Dynamic Contrast Enhanced Prostate MRI Synthesis.\\
}

\author{\IEEEauthorblockN{Divya Bharti$^1{*}\quad$ Sriprabha Ramanarayanan$^{1,2}\quad$ Sadhana S$^1\quad$ Kishore Kumar M$^1\quad$ Keerthi Ram$^2\quad$ \\
Harsh Agarwal$^3\quad$ Ramesh Venkatesan$^3\quad$ Mohanasankar Sivaprakasam$^{1,2}\quad$}
\and 
\hspace{50mm}\textit{$^1$Indian Institute of Technology Madras (IITM), India}\\
\hspace{50mm}\textit{$^2$Healthcare Technology Innovation Centre (HTIC), India}\\
\hspace{50mm}\textit{$^3$GE HealthCare (GE), India}\\

\hspace{50mm}{\tt\small${*}$b13.divya@gmail.com}}

\maketitle

\begin{abstract}
Dynamic Contrast-Enhanced Magnetic Resonance Imaging (DCE-MRI) is a medical imaging technique that plays a crucial role in the detailed visualization and identification of tissue perfusion in abnormal lesions and radiological suggestions for biopsy. However, DCE-MRI involves the administration of a Gadolinium-based (Gad) contrast agent, which is associated with a risk of toxicity in the body. Previous deep learning approaches that synthesize DCE-MR images employ unimodal non-contrast or low-dose contrast MRI images lacking focus on the local perfusion information within the anatomy of interest. We propose AAD-DCE, a generative adversarial network (GAN) with an aggregated attention discriminator module consisting of global and local discriminators. The discriminators provide a spatial embedded attention map to drive the generator to synthesize early and late response DCE-MRI images. Our method employs multimodal inputs - T2 weighted (T2W), Apparent Diffusion Coefficient (ADC), and T1 pre-contrast for image synthesis. Extensive comparative and ablation studies on the ProstateX dataset show that our model (i) is agnostic to various generator benchmarks and (ii) outperforms other DCE-MRI synthesis approaches with improvement margins of +0.64 dB PSNR, +0.0518 SSIM, -0.015 MAE for early response and +0.1 dB PSNR, +0.0424 SSIM, -0.021 MAE for late response, and (ii) emphasize the importance of attention ensembling. Our code is available at https://github.com/bhartidivya/AAD-DCE.
\end{abstract}

\vspace{-3mm}
\section{Introduction}
Dynamic contrast-enhanced Magnetic Resonance Imaging (DCE-MRI) is a medical image scanning technique that quantifies tumor vasculature and perfusion characteristics \cite{b1}. The angiogenesis is captured by injecting a Gadolinium (Gad)-based contrast agent. This contrast agent exhibits a notable characteristic of rapid wash-in and wash-out kinetics occurring within a few seconds of Gad injection in the suspicious tissues, distinguishing it from normal healthy tissues. DCE-MRI plays a crucial role in prostate imaging, where the radiologists use a PI-RADS\footnote{https://www.acr.org/-/media/ACR/Files/RADS/Pi-RADS/PIRADS-v2-1.pdf} score (in the scale 1 to 5) to assess the cancer severity by visualizing non-contrast images (T2 or diffusion-weighted images (DWI)). A score of 3 suggests DCE-MRI acquisition, which helps to reduce unnecessary biopsies by 25\% \cite{b10}, avoiding over-diagnosis of insignificant cancers. However, Gad-based contrast agents cause patient discomfort and other contraindications like Nephrogenic Systemic Fibrosis \cite{b2}. Therefore, there is a need to decrease the dosage or refrain from using a Gad-based contrast agent in DCE-MRI.


\setlength{\belowcaptionskip}{-5mm}
\begin{figure}[t]
\vspace{-2mm}
    \centering
    \includegraphics[width=0.9\linewidth]{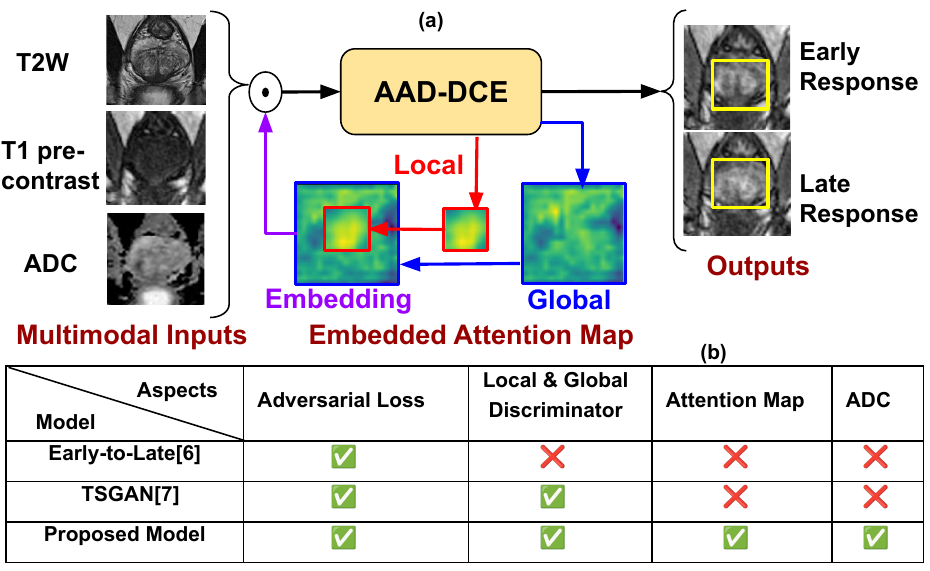}
    \caption{(a) Concept diagram of AAD-DCE. Multimodal non-contrast inputs, with aggregated attention maps, synthesize early and late response DCE-MRI. (b) Previous DCE-MRI synthesis methods use adversarial training without focusing on key anatomical regions. The proposed approach integrates a discriminator-based global and local attention mechanism to enhance the generator's performance.}
    \label{fig:block diagram}
\end{figure}

Deep learning methods have been explored for MRI reconstruction tasks (\cite{20},\cite{21}) and to reduce or eliminate the use of Gad-based DCE-MRI. The Generative Adversarial Network (GAN) \cite{b3} using a 3D U-Net-like generator produces 3D isotropic contrast-enhanced images from a 2D T2-flair image stack by adopting spatial pyramid pooling, enhanced residual blocks, and deep supervision. Retina U-Net \cite{b4} extracts the semantic features from non-contrast brain MR images and uses a synthesis module for contrast-enhanced images. The convolutional neural network (CNN) based network \cite{b5} generates full-dose late-response images from pre-contrast and low-contrast images for brain MRI. A U-Net-based conditional-GAN \cite{b6} with residual loss function synthesizes late-response images from early-response breast MRI images. DCE-Diff \cite{b16} is a diffusion-based generative model that maps non-contrast to contrast-enhanced prostate images.

\begin{figure*}
    \centering
     {\includegraphics[width=0.92\linewidth]{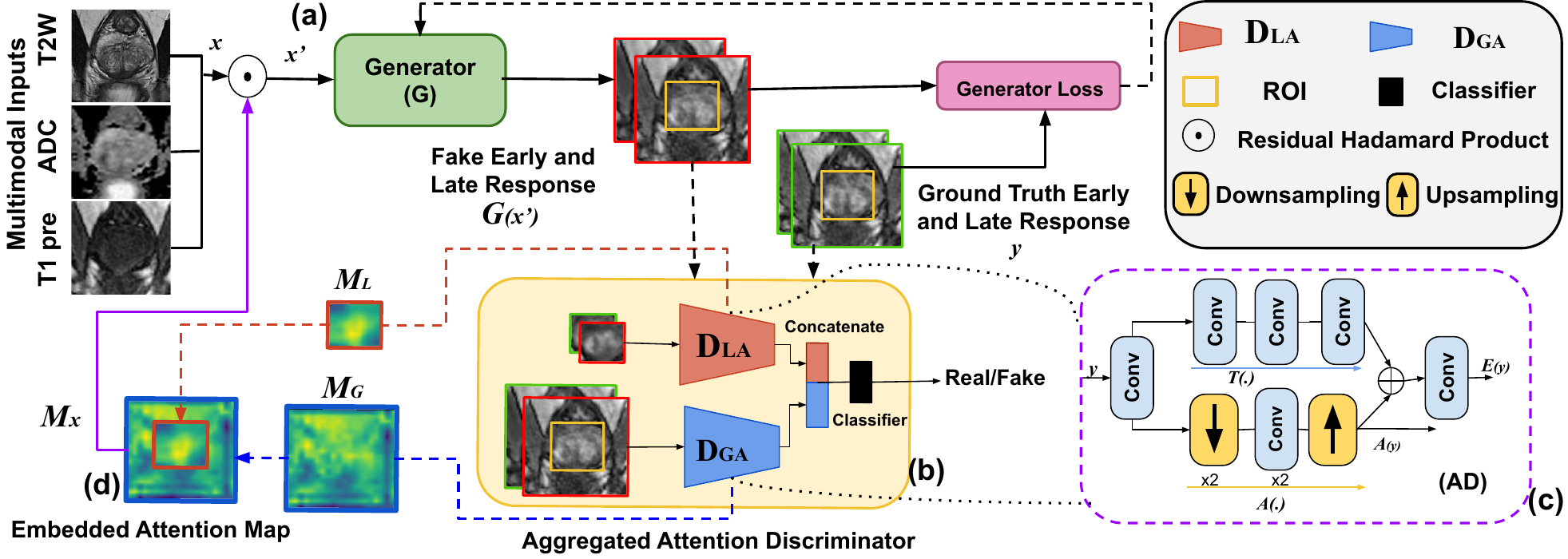}}
    \caption{(a) AAD-DCE architecture with a generator and an Aggregated Attention Discriminator (AAD) module. (b) AAD  with local and global attention discriminators, $D_{LA}$ and $D_{GA}$, respectively, utilizes (c) Attention Discriminator (AD) architecture. (d) Embedded attention map $M_{x}$.}
    \label{fig:network architecture}
\end{figure*}

However, the former methods (\cite{b3},\cite{b4},\cite{b5},\cite{b6}) either depend on low-dose images or do not completely utilize the perfusion information from the non-contrast DCE-MRI data. In DCE-MRI, the Apparent Diffusion Coefficient (ADC) map computed using DWI carries useful information needed for generating the perfusion information. Secondly, the GAN-based methods adopt a global discriminator that is not specialized to consider the importance of local distribution related to the perfusion information within the anatomy of interest. DCE-Diff focuses on structural correlation without local context and suffers from higher computation costs for training and inference.

The goal of this work is to synthesize DCE-MRI early and late response images leveraging the complementary information from multimodal non-contrast MRI inputs and provide nuanced and detailed feature representation for perfusion (Fig. \ref{fig:block diagram}). Our method employs a GAN framework where the generator obtains additional guidance via an attention map computed and fed back from its discriminator. The attention map helps to focus more on the most discriminative areas between abnormal and healthy regions in the anatomy under study. Our method uses two discriminators to learn (i) the global structural correlation between the pre-contrast MRI and non-contrast T2 Weighted (T2W) MR images and (ii) local perfusion information using the ADC maps. In contrast, previous methods (TSGAN\cite{b7}, ReconGLGAN\cite{b8}) use two discriminators for optimization without attention guidance.

We propose an Aggregated Attention-based discriminator (AAD) module consisting of two discriminators for global and local processing of the region of interest (ROI). The discriminators provide embedded spatial attention maps, each highlighting perfusion at fine and abstract levels. The two maps are aggregated and the ensembled map, rich in local details, guides the generator for improved synthesis of post-contrast images. Different from the previous attention-based discriminator \cite{b9}, the proposed Aggregated Attention-based discriminator introduces a composition of attention maps to drive the generator.
We summarize our contributions as follows:
1) An end-to-end trainable GAN for DCE-MR image synthesis from non-contrast multimodal MRI inputs, namely T2W, ADC, and T1 pre-contrast, with aggregated attention guidance to the generator to learn discriminative features that highlight hyper-intense abnormal regions in the prostate anatomy.
2) A dual-discriminator framework with aggregated attention modules that (i) enables adversarial learning to capture both global and local perfusion characteristics of the prostate anatomy and (ii) learns an aggregated attention map that highlights essential details in the region of interest to guide the generator. Our discriminator is agnostic to varying generator architectures.
3) Extensive experiments on the ProstateX dataset demonstrate that the proposed model predicts early and late response contrast-enhanced images with improvement margins (i)+0.64 dB PSNR, +0.0518 SSIM, -0.015 MAE for early response, and (ii) +0.1 dB PSNR, +0.0424 SSIM, -0.021 MAE for late response, surpassing the second best performing model, DCE-Diff. Our experiments highlight the importance of aggregated attention and ADC for accurate DCE-MRI synthesis.

\section{Proposed Method}
Our method is based on GAN, where a generator (G) and a discriminator (D) compete with each other to synthesize ground truth-like images from multi-modal input images. The core concept of AAD-DCE is to employ embedded spatial attention maps generated by the Aggregated Attention Discriminator (AAD) to guide the generator. This approach drives the generator to focus more on ROI, thus enhancing the post-contrast images. The input $\boldsymbol{x}$ consists of T2W, ADC, and T1 pre-contrast images concatenated along the channel axis while $\boldsymbol{y}$ represents the early and late response DCE-MR images as seen in Fig. \ref{fig:network architecture}(a).

\begin{table*}[t]
\centering
\caption{Quantitative comparison of the generated early and late response DCE-MRI images between AAD-DCE and other models.}
\vspace{-6mm}
\label{tab:quantitative results}
\begin{tabular}{cccclccccc}
                                &                                      &                                     & \multicolumn{1}{l}{}            &            &                           &                                      &                                     &                          &                           \\ \hline
\multirow{2}{*}{\textbf{Model}} & \multicolumn{4}{c}{\textbf{Early Response}}                                                                               &                           & \multicolumn{3}{c}{\textbf{Late Response}}                                                            &                           \\ \cline{2-10} 
                                & \textbf{PSNR$\uparrow$}              & \textbf{SSIM$\uparrow$}             & \multicolumn{2}{c}{\textbf{MAE$\downarrow$}} & \textbf{FID $\downarrow$} & \textbf{PSNR$\uparrow$}              & \textbf{SSIM$\uparrow$}             & \textbf{MAE$\downarrow$} & \textbf{FID $\downarrow$} \\ \hline
ConvLSTM                        & 14.92 $\pm$ 1.50                     & 0.2397 $\pm$ 0.04                   & \multicolumn{2}{c}{0.139}                   & 118.706                   & 15.27 $\pm$ 2.56                     & 0.2392 $\pm$ 0.06                   & 0.135                    & 115.480                   \\
CycleGAN                        & 18.61 $\pm$ 1.90                     & 0.5134 $\pm$0.06                    & \multicolumn{2}{c}{0.128}                   & 40.4371                   & 17.20 $\pm$ 2.40                     & 0.4982 $\pm$ 0.06                   & 0.129                    & 41.6501                   \\
Pix2Pix                         & 19.53 $\pm$ 1.97                     & 0.5719 $\pm$ 0.05                   & \multicolumn{2}{c}{0.066}                   & 25.9182                   & 19.35 $\pm$ 1.94                     & 0.5546 $\pm$ 0.06                   & 0.128                    & 27.0816                   \\
RegGAN                          & 20.56 $\pm$ 0.02                     & 0.5966 $\pm$ 0.02                   & \multicolumn{2}{c}{0.057}                   & 23.7963                   & 19.89 $\pm$ 0.02                     & 0.5803 $\pm$ 0.02                   & 0.069                    & 22.6124                   \\
TSGAN                           & 21.16 $\pm$ 3.50                     & 0.6253 $\pm$ 0.10                   & \multicolumn{2}{c}{0.069}                   & 23.7533                   & 20.08 $\pm$ 2.64                     & 0.5926 $\pm$ 0.09                   & 0.074                    & 24.6665                   \\
ResViT                          & 20.12 $\pm$ 1.61                     & 0.6308 $\pm$ 0.05                   & \multicolumn{2}{c}{0.063}                   & 32.4624                   & 19.60 $\pm$ 1.69                     & 0.6153 $\pm$ 0.05                   & 0.070                    & 30.0611                   \\
\multicolumn{1}{l}{DCE-Diff}    & \multicolumn{1}{l}{22.10 $\pm$ 1.79} & \multicolumn{1}{l}{0.6700 $\pm$ 0.05} & \multicolumn{2}{c}{0.040}                   & \textbf{10.5900}            & \multicolumn{1}{l}{21.73 $\pm$ 1,95} & \multicolumn{1}{l}{0.6500 $\pm$ 0.06} & 0.050                     & \textbf{7.2600}             \\
\textbf{AAD-DCE}                   & \textbf{22.74 $\pm$ 1.94}            & \textbf{0.7218 $\pm$ 0.05}          & \multicolumn{2}{c}{\textbf{0.025}}          & 18.5352                   & \textbf{21.83 $\pm$ 2.15}            & \textbf{0.6924 $\pm$ 0.06}          & \textbf{0.029}          & 19.8306                   \\ \hline
\end{tabular}
\end{table*}

\subsection{Generator}\label{AA}



The Generator (G) can be any image-to-image mapping CNN or transformer-based network. Our discriminator is adaptable to various generator architectures (Section III B). The generator architecture in Fig. \ref{fig:network architecture}(a), is an encoder-decoder architecture with residual blocks.


\subsection{Aggregated Attention Discriminator (AAD)}
The architecture of the AAD block is shown in Fig. \ref{fig:network architecture}(b). It consists of a Global Attention discriminator $D_{GA}$, a Local Attention discriminator $D_{LA}$, and a Classifier ($\Psi_C$). $D_{GA}$ and $D_{LA}$ are global and local feature extractors that employ a spatial guidance module named Attention Discriminator (AD). The input to $D_{GA}$ is the whole image ($H \times W$) whereas $D_{LA}$ operates only the ROI ($H^{\prime} \times W^{\prime}$).
The AD in Fig. \ref{fig:network architecture}(c) comprises two key components: the Attention branch and the Trunk branch inspired by RAM \cite{b19}. The Trunk branch made up of convolutional layers, extracts low-level features from the input $\boldsymbol{y}$ to produce the output T($\boldsymbol{y}$). Note that our trunk branch is much simpler than in RAM \cite{b19} while maintaining feature extraction capability. The Attention branch, using the bottom-up top-down structure \cite{b18}, learns an attention map A($\boldsymbol{y}$), which modulates the trunk branch's output by applying weights. The output of $D_{GA}$ and $D_{LA}$ that utilizes the AD module are shown in (\ref{equ:2}) and (\ref{equ:3}),


\vspace{-2mm}
\begin{equation}
\vspace{-2mm}
E_{G}= \left(A_{G}(y)+1\right) \times T_{G}(y)  
\label{equ:2}
\end{equation}
\begin{equation}
E_{L}= \left(A_{L}(y)+1\right) \times T_{L}(y)  
\label{equ:3}
\end{equation} 
where $A_{G}(y)$, $A_{L}(y)$ are the attention maps from attention branch and $T_{G}(y)$, $T_{L}(y)$ are trunk branch outputs for $D_{GA}$ and $D_{LA}$ respectively.
The $N_{C}$ dimensional feature vectors $E_{G}$ and $E_{L}$ are concatenated into a $2N_{C}$ dimensional vector and passed to the classifier ($\Psi_C$) as shown in (\ref{equ:4}). This vector is then fed into a fully connected layer, followed by a sigmoid activation function to classify real or fake.
\vspace{-2mm}
\begin{equation}
D({y})=\Psi_C\left(E_{G}\|\ E_{L}\right)
\label{equ:4}
\end{equation}
The mean value of the attention maps $M_{G}$ and $M_{L}$ obtained from the two discriminators, $D_{GA}$ and $D_{LA}$  are aggregated by embedding $M_{L}$ into $M_{G}$ to give $M_{x}$. $M_{x}$ is infused into the multimodal inputs ($x$) using Residual Hadamard Product (RHP) as seen in (\ref{equ:5}) and fed to the  Generator.
\vspace{-2mm}
\begin{equation}
    x^{\prime}=\left(M_{x} +1\right) \times x 
\label{equ:5}
\end{equation}
As the infusion of attention map is based on RHP, it is initialized to one at the start of the training. 
The adversarial GAN loss, generator loss, and the final objective function are: 
\vspace{-1mm}
\begin{equation}
    \begin{split}
        L_{G A N}(G, D)= & \mathbb{E}_{y \sim \mathcal{Y}}[\log D(y)]+ \\
 & \mathbb{E}_{x \sim \mathcal{X}}\left[\log \left(1-D\left(G\left(x^{\prime}\right)\right)\right)\right] 
    \end{split}
\label{equ:6}
\end{equation}
\vspace{-2mm}
\begin{equation}
    L_{L 1}(G)=\mathbb{E}_{x, y}\left[\left\|y-G\left(x^{\prime}\right)\right\|_1\right]
\label{equ:7}
\end{equation}
\vspace{-2mm}
\begin{equation}
    \arg \min _G \max _D L_{G A N}(G, D)+\lambda L_{L 1}(G)
\label{equ:8}
\end{equation}

\begin{figure*}[!t]
    \centering
     {\includegraphics[width=0.96\linewidth]{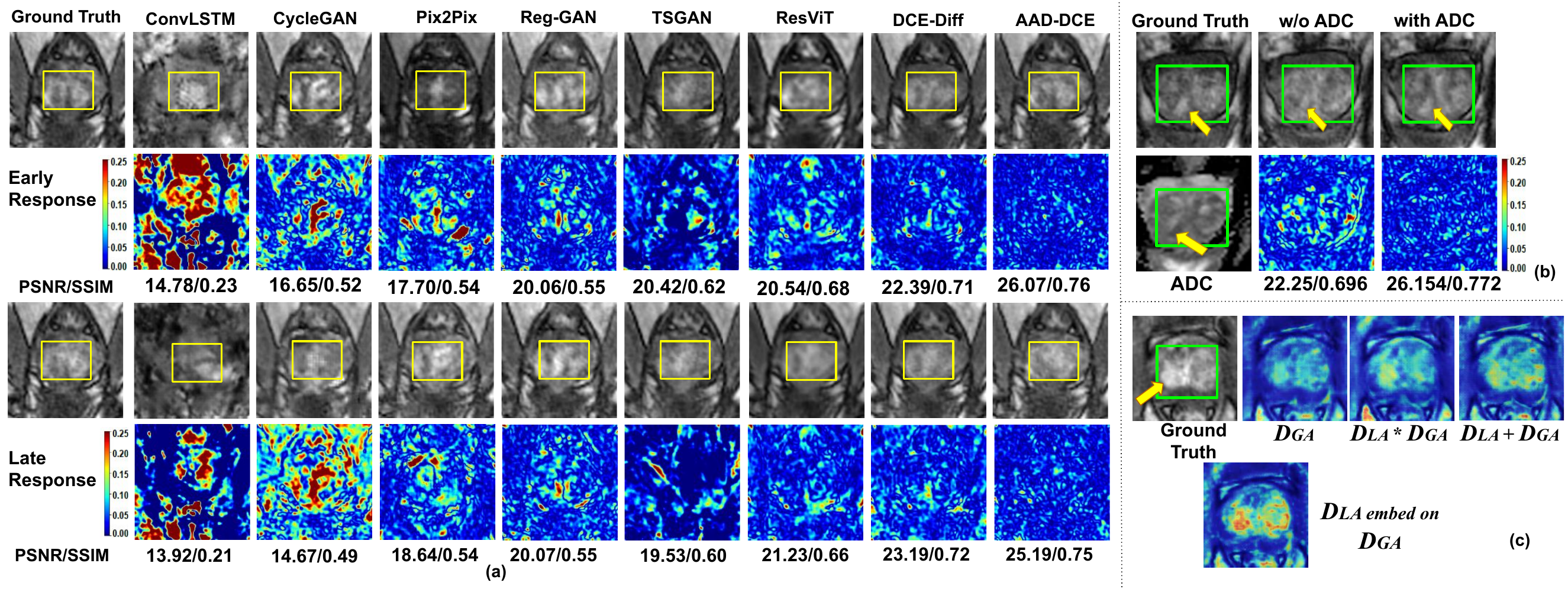}}
    \caption{(a) Visual results of early and late response for the ProstateX dataset with error maps. The yellow bounding box marks the ROI. (b) Ablative study with and without ADC maps. (c) Attention maps from different ensembling methods. Attention embedding enables better focus on suspicious regions.}
    \label{fig:image_posxl}
\end{figure*}

\section{Experiments and Results}
\subsection{Dataset and Implementation Details}
We trained our model on the open-source ProstateX dataset \cite{b17} consisting of T2-Weighted, ADC, T1 pre-contrast, and DCE images. The dataset consists of 346 patient studies with 5520 images (4410 for training and 1104 for validation). In the DCE data, the early enhancement usually occurs within 10 seconds of the appearance of the injected contrast agent, and therefore, the early and late response time points are selected accordingly. The data is registered using SimpleITK rigid registration and is resampled to $H \times W \times 16$. Here $H \times W$ is $160\times160$ and $H^\prime \times W^\prime$ is $60\times60$. We set $N_{C}$ as 64.
\begin{table}[h]
\centering
\caption{Importance of ADC}
\vspace{-2mm}
\label{tab:adc_ablative}
\begin{tabular}{ccccc}
\hline
\textbf{DCE-MRI} & \multicolumn{2}{c}{\textbf{Early Response}} & \multicolumn{2}{c}{\textbf{Late Response}} \\ \hline
\textbf{ADC}     & w/o ADC          & w/ ADC                   & w/o ADC          & w/ ADC                  \\ \hline
\textbf{PSNR}    & 21.50            & \textbf{22.74}           & 20.43            & \textbf{20.64}          \\ \hline
\textbf{SSIM}    & 0.6841           & \textbf{0.7218}          & 0.6404           & \textbf{0.6924}         \\ \hline
\end{tabular}
\end{table}

The models are trained using Pytorch 2.0 on a 24GB RTX 3090 GPU. The Adam optimizer ($\beta_{1}=0.9$, $\beta_{2}=0.999$, learning rate 1e-3) is used for 200 epochs, batch size of 4, and $\lambda$ is 10. The evaluation metrics are Peak Signal-to-Nosie Ratio (PSNR), Structural Similarity Index (SSIM), Mean Absolute Error (MAE), and Frechet Inception Distance (FID).

\subsection{Results and Discussion}
\textbf{1. Comparative study with previous methods:} We compare AAD-DCE with baseline models, Pix2Pix \cite{b11}, CycleGAN \cite{b12}, ConvLSTM \cite{b13}, Reg-GAN \cite{b14}, ResViT \cite{b15}, TSGAN \cite{b7}, and DCE-Diff \cite{b16}. 
The quantitative results in Table \ref{tab:quantitative results} show that our model outperforms other GAN-based, transformer-based, and convLSTM methods in most cases. Our model outperforms DCE-Diff, by +0.64dB, +0.1 dB in PSNR, +0.0518, +0.0424 in SSIM, and -0.015, -0.021 in MAE for early and late response respectively, except in terms of FID due to diffusion models' ability to learn image distribution through probabilistic framework. Fig. \ref{fig:image_posxl}(a) illustrates the synthesis results highlighting the abnormal regions in the transition and peripheral zones (TZ and PZ) of the prostate. We observe that AAD-DCE is most comparable to the ground truth with fewer errors in the error map and better preserves the global and local details of the post-contrast images.
We believe that: (i) While $D_{GA}$ focuses on the overall anatomical features, $D_{LA}$ targets the suspicious regions within the prostate via the attention information. (ii) ADC provides perfusion information, while T2W and T1 pre-contrast MR images establish the structural correlations. Integrating these modalities enables learning the complementary information, resulting in more accurate and detailed structural representations and perfusion information in the predicted images.
\setlength{\tabcolsep}{4pt}
\begin{center}
\begin{table}[]
\centering
\caption{Ablative study on attention ensembling}
\vspace{-2mm}
\label{tab:discriminator_abalative}
\begin{tabular}{lcccc}
\hline
\textbf{Attention Maps} & $D_{GA}$ & $D_{LA}$*$D_{GA}$ & $D_{LA}+D_{GA}$ & \begin{tabular}[c]{@{}c@{}}$D_{LA}$ embed\\  on $D_{GA}$\end{tabular} \\ \hline
\textbf{PSNR}       & 22.06  & 21.66           & 22.38         & \textbf{22.73}                                                      \\ \hline
\textbf{SSIM}       & 0.7012   & 0.7014            & 0.7038          & \textbf{0.7218}                                                       \\ \hline
\end{tabular}
\end{table}
\end{center}
\vspace{-5mm}

\vspace{-3.5mm}
\textbf{2. Importance of ADC:}\label{AAA}
Table \ref{tab:adc_ablative} shows the importance of using ADC as inputs. The incorporation of ADC images containing perfusion information improves the post-contrast image synthesis. Fig. \ref{fig:image_posxl}(b) shows the abnormality in PZ that correlates with the tumor region findings in the ground truth (highlighted in yellow arrows).

\textbf{3. Ablative study on attention ensembling:} We study the effect of only the global attention and different ways of ensembling the global and local attention maps, namely additive, multiplicative, and embedding operations (Table \ref{tab:discriminator_abalative}). In Fig. \ref{fig:image_posxl}(c), the embedded attention map focuses more precisely on the regions with contrast uptake
with greater clarity. 

\begin{figure}
    \centering
    \includegraphics[width=0.8\linewidth]{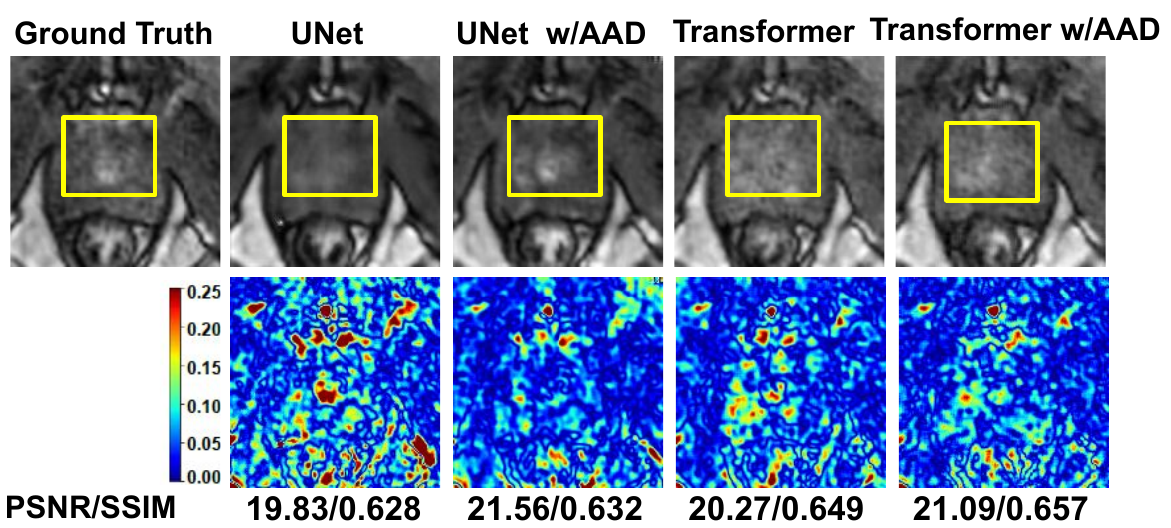}
    \vspace{-2mm}
    \caption{Generator architecture with AAD enhances focus on the ROI.}
    \label{fig:gen_ablative}
\end{figure}

\textbf{4. Ablative study on various generator architectures:} We have evaluated the proposed AAD on various generator benchmarks - encoder-decoder CNN (U-Net) \cite{b11}, vision transformer \cite{b15}. A U-Net with AAD improves PSNR by +1.49 dB and SSIM by +0.0387, while for the transformer-based generator, AAD provides improvements of +1.77 dB in PSNR and +0.0432 in SSIM (Fig. \ref{fig:gen_ablative}).

\textbf{5. Model Parameters}: Comparing with the recent state-of-the-art, DCE-Diff, shows that our model has 19.93 million parameters with superior performance over DCE-Diff with 125.08 million parameters. 


\section{Conclusion}
This work presents a GAN framework with aggregated attention feedback guidance to the generator from the discriminator to synthesize DCE-MR images from multimodal non-contrast MRI inputs. Extensive experimentation with comparative models and ablation studies with ADC, attention ensembling methods, and various generator architectures show that the proposed method can synthesize higher-quality DCE-MR images. We are currently aiming at clinical validation studies for practical use.

\end{document}